%% file: main.tex
\documentclass[10pt, letter, twoside, titlepage]{article} 

\usepackage[letterpaper, portrait, margin=3cm]{geometry}
\usepackage[document]{ragged2e}
\usepackage{url,hyperref,lineno,microtype}
\usepackage[singlespacing]{setspace}
\modulolinenumbers[5]

\usepackage{natbib}
\setcitestyle{authoryear,open={(},close={)}}

\usepackage{adjustbox}
\usepackage{color,soul}

\begin{document}
\onecolumn
\linespread{1.0}
\fontfamily{ptm}\selectfont

\setcounter{equation}{0}
\setcounter{figure}{0}
\setcounter{section}{0}
\clearpage
\pagenumbering{arabic}

\Large
 \begin{center}
\textbf{Physiological Tremor Increases when Skeletal Muscle is Shortened: Implications for Fusimotor Control}\\ 

\large
Kian Jalaleddini\,$^{1,\S}$, Akira Nagamori\,$^{1,\S}$, Christopher M. Laine\,$^{1}$, Mahsa A. Golkar\,$^{2}$, Robert E. Kearney\,$^{2}$, Francisco J Valero-Cuevas
$^{1,3,\ddag}$ \\

\end{center}
\small  
$^{1}$Division of Biokinesiology and Physical Therapy, University of Southern California, USA \\
$^{2}$Department of Biomedical Engineering, McGill University, Canada \\
$^{3}$Department of Biomedical Engineering, University of Southern California, USA\\
$^\S$ Equal Contribution\\
$^\ddag$ Corresponding Author\\

\justify
\normalsize

\input{summary.tex}
\section{Abstract}
The involuntary force fluctuations associated with physiological (as distinct from pathological) tremor are an unavoidable component of human motor control. While the origins of physiological tremor are known to depend on muscle afferentation, it is possible that the mechanical properties of muscle-tendon systems also affect its generation, amplification and maintenance. In this paper, we investigated the dependence of physiological tremor on muscle length in healthy individuals. We measured physiological tremor during tonic, isometric plantarflexion torque at 30\% of maximum at three ankle angles. The amplitude of physiological tremor increased as calf muscles shortened in contrast to the stretch reflex whose amplitude decreases as muscle shortens. We used a published closed-loop simulation model of afferented muscle to explore the mechanisms responsible for this behavior. We demonstrate that changing muscle lengths does not suffice to explain our experimental findings. Rather, the model consistently required the modulation of $\gamma$-static fusimotor drive to produce increases in physiological tremor with muscle shortening---while successfully replicating the concomitant reduction in stretch reflex amplitude. This need to control $\gamma$-static fusimotor drive explicitly as a function of muscle length has important implications. First, it permits the amplitudes of physiological tremor and stretch reflex to be decoupled. Second, it postulates neuromechanical interactions that require length-dependent $\gamma$ drive modulation to be independent from $\alpha$ drive to the parent muscle. Lastly, it suggests that physiological tremor can be used as a simple, non-invasive measure of the afferent mechanisms underlying healthy motor function, and their disruption in neurological conditions.


\newpage
\input{introduction.tex}
\input{methods.tex}
\input{results.tex}
\input{discussion.tex}

\input{acknowledgement.tex}
\bibliography{reference}
\end{document}

%% file: summary.tex
\section{Key Point Summary}
\begin{itemize}
\item In tonic, isometric, plantarflexion contractions, physiological tremor increases as the ankle joint becomes plantarflexed.
\item Modulation of physiological tremor as a function of muscle stretch differs from that of the stretch reflex amplitude.
\item Amplitude of physiological tremor may be altered as a function of reflex pathway gains.
\item Healthy humans likely increase their $\gamma$-static fusimotor drive when muscles shorten.
\item Quantification of physiological tremor by manipulation of joint angle may be a useful experimental probe of afferent gains and/or the integrity of automatic fusimotor control.
\end{itemize}

%% file: introduction.tex
\section{Introduction}
Physiological tremor is the unavoidable tendency of muscles to generate involuntary, rhythmic oscillations in the frequency range of 5-12 Hz \citep{lippold1971physiological,sowman2005methods}. There is an emerging consensus that muscle afferentation and the integrity of the spinal cord stretch reflex circuitry lie at the heart of physiological tremor \citep{lippold1971physiological,young1980physiological,christakos2006parallel,laine2016dynamics,cresswell2000significance}. Given the cardinal role that disrupted muscle afferentation plays in the clinical presentation of neurological conditions, it is critical to develop noninvasive means to assess reflex function through measurements of physiological tremor amplitude. One possibility is to leverage the known increase in stretch reflex amplitudes with muscle stretch, potentially due to increased firing of muscle spindles \citep{mirbagheri2000intrinsic,mileusnic2006spindle}. Dependence of physiological tremor on muscle stretch has not been characterized. Doing so might enable noninvasive methods to assess the integrity of spinal circuitry without the need for electrical or mechanical perturbations.

Spindle primary afferents are sufficiently sensitive to respond to the changes in muscle length induced by physiological tremor in contracting muscles \cite{hagbarth1979participation}. Since stretch reflex is often used to assess the sensitivity of spindle afferents, we hypothesized, therefore, that the amplitude of physiological tremor would increase with muscle stretch provided muscle activation was held constant. The human ankle plantarflexor muscle is a well established model to study muscle and joint neuromechanics \citep{mugge2010rigorous,de2010relation}, stretch reflex responses \citep{kearney1997identification,alibiglou2008relation,toft1991mechanical}, afferent gains (e.g., H reflex) \citep{capaday1986amplitude}, and physiological tremor \citep{laine2014task}. Consequently, we investigated physiological tremor in healthy adults as they produced tonic isometric plantarflexion ankle torque at different joint angles. To elucidate the mechanical and neural factors that could explain our results, we also simulated a closed-loop model of afferented muscle as in \citep{laine2016dynamics}, scaled to the architecture of the gastrocnemius .

Our findings show that the amplitude of physiological tremor is related to (and can be manipulated through) changes in muscle length. Further, our simulations establish that it is not sufficient to simply alter muscle mechanics, but adjustment of spindle sensitivity is necessary to replicate the changes in the physiological tremor associated with muscle stretch. Given that a physiologically-plausible fusimotor mechanism could explain our experimental results, we discuss the implications of this for the development of simple and powerful techniques for assessing the integrity of reflex pathways in health and disease.

%% file: methods.tex
\section{Methods}
\subsection{Experimental Methods}
Seven subjects with no history of neuromuscular disease were recruited (Age of 32 $\pm$ 2.2, 2 female). Subjects gave their informed consent prior to participation, and all procedures were approved by the McGill University Institutional Review Board. 

\subsubsection{Experimental Setup}
\begin{figure}
\begin{center}
     \includegraphics [scale=1.2]{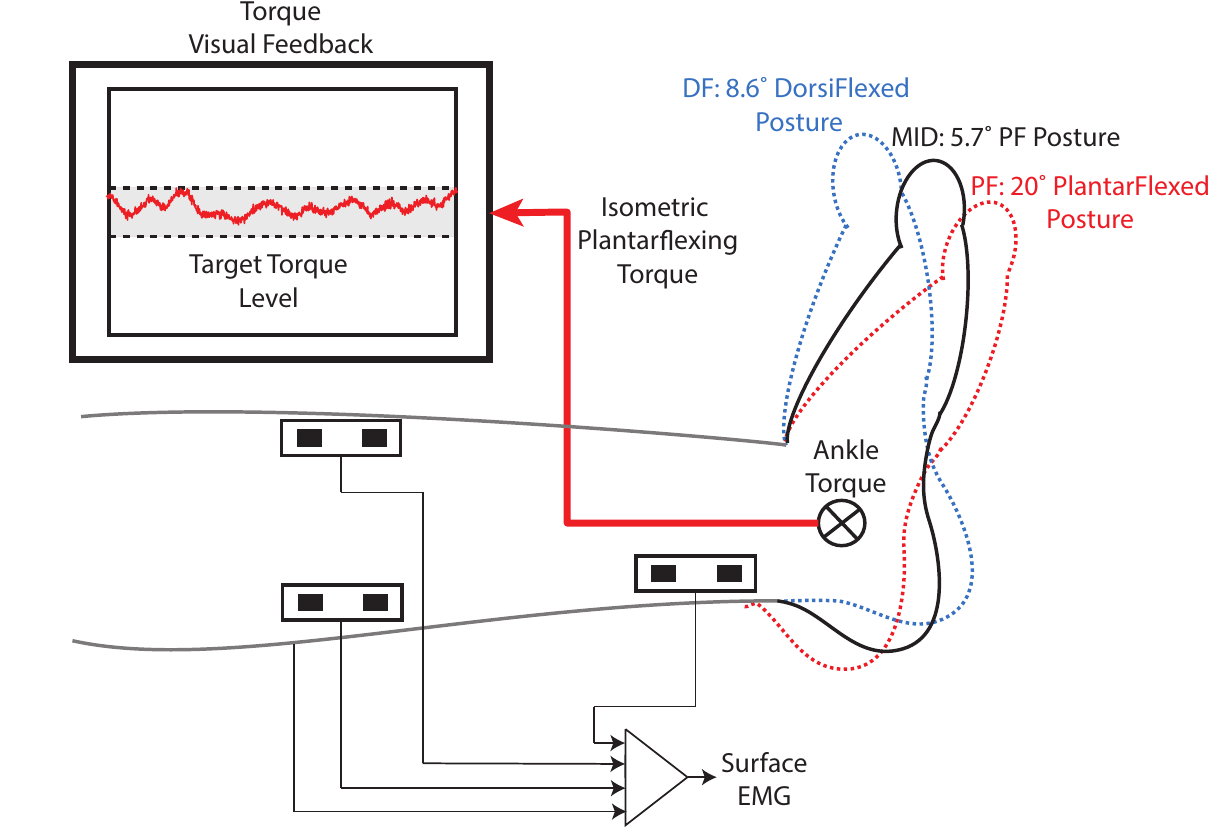}
     \caption{Subjects lay supine with the knee extended and were instructed to produce a constant isometric plantarflexing ankle torque of 30\% of their maximum voluntary contraction, aided by an overhead monitor. Three joint angles were tested: $20 ^{\circ}$ plantarFlexed (PF), $5.7 ^{\circ}$ PF from neutral (MID), and $8.6 ^{\circ}$ dorsiflexed (DF) from neutral where the neutral angle was defined as the 90 $^\circ$ angle between the shank and foot. Ankle torque and EMG from the main plantarflexors and dorsiflexor were recorded.}
     \label{experimentalSetup}
\end{center}
\end{figure}

The subjects lay supine while the left foot was placed inside a custom-made fiberglass boot rigidly attached to a hydraulic actuator (Figure~\ref{experimentalSetup}). The boot constrained the whole limb movement except the ankle movement in the sagittal plane. The actuator was programmed to maintain a particular ankle angle.

EMG signals were recorded from the main plantarflexor and dorsiflexor muscles using standard, active, single differential surface electrodes supplied with the Bagnoli System (Delsys, Inc., Boston). EMGs were recorded from the medial and lateral heads of the gastrocnemius, soleus and tibialis anterior muscles. The reference electrode was a DermaSport electrode (American Imex, Irvine) placed on the left knee joint. The EMG signals were amplified 1000 times and high-pass filtered at 20 Hz prior to acquisition.

The ankle angle was measured using a 6273 BI Technologies potentiometer (TT Electronics, Woking, UK). The ankle torque was measured using a 2110-5K Lebow transducer (Honeywell, Columbus). All signals were low-pass filtered at 486.3 Hz and digitized at 1 kHz with 24-bit resolution using a set of NI-4472 A/D cards (National Instruments, Austin).

\subsubsection{Experimental Trials}
The experiment consisted of three types of trials: (I) \emph{Maximum Voluntary Contraction} (MVC) trials. Subjects were instructed to perform five maximum voluntary contractions separated by 10 s rest intervals. The MVC was taken as the maximum torque in this trial. (II) \emph{Passive torque} trials of 10 s. At each joint angle, subjects were instructed to relax and the passive torque was taken as the average of the torque. (III) \emph{Tonic Contraction} trials of 30 s. Each subject was instructed to maintain a tonic voluntary contraction of 30\% of MVC torque aided by a visual feedback of their filtered torque in real-time with an eighth-order, low-pass, Bessel filter with 0.7 Hz cutoff.

The hydraulic actuator moved the ankle to the angle randomly selected from: $20 ^{\circ}$ PlantarFlexed (PF), $5.7 ^{\circ}$ plantarflexed from neutral (MID), and $8.6 ^{\circ}$ DorsiFlexed (DF) from neutral where the neutral angle was defined as the 90 $^\circ$ angle between the shank and foot. At each angle, the subjects performed one MVC trial, one passive-torque trial and three tonic contraction trials successively. One minute of rest was imposed between trials to prevent muscle fatigue.

\subsubsection{Data Analysis}
To describe the tremor in torque during each trial, we conducted a spectral analysis.  First, the tonic contraction torque was normalized by subtracting the passive torque and dividing by the MVC torque at each angle. Second, the contribution of slow-varying torque was removed using a high-pass, sixth order, butterworth filter with 5 Hz cut-off frequency. Third, the power spectral density of the torque was calculated using 1 s Gaussian-tapered windows. The power of the physiological tremor was calculated as the average of the power spectral density in the 5-12 Hz band. Lastly, the power of the physiological tremor was normalized to that of the DF posture for analysis of the group results.

\begin{figure}
\begin{center}
     \includegraphics [scale=0.7]{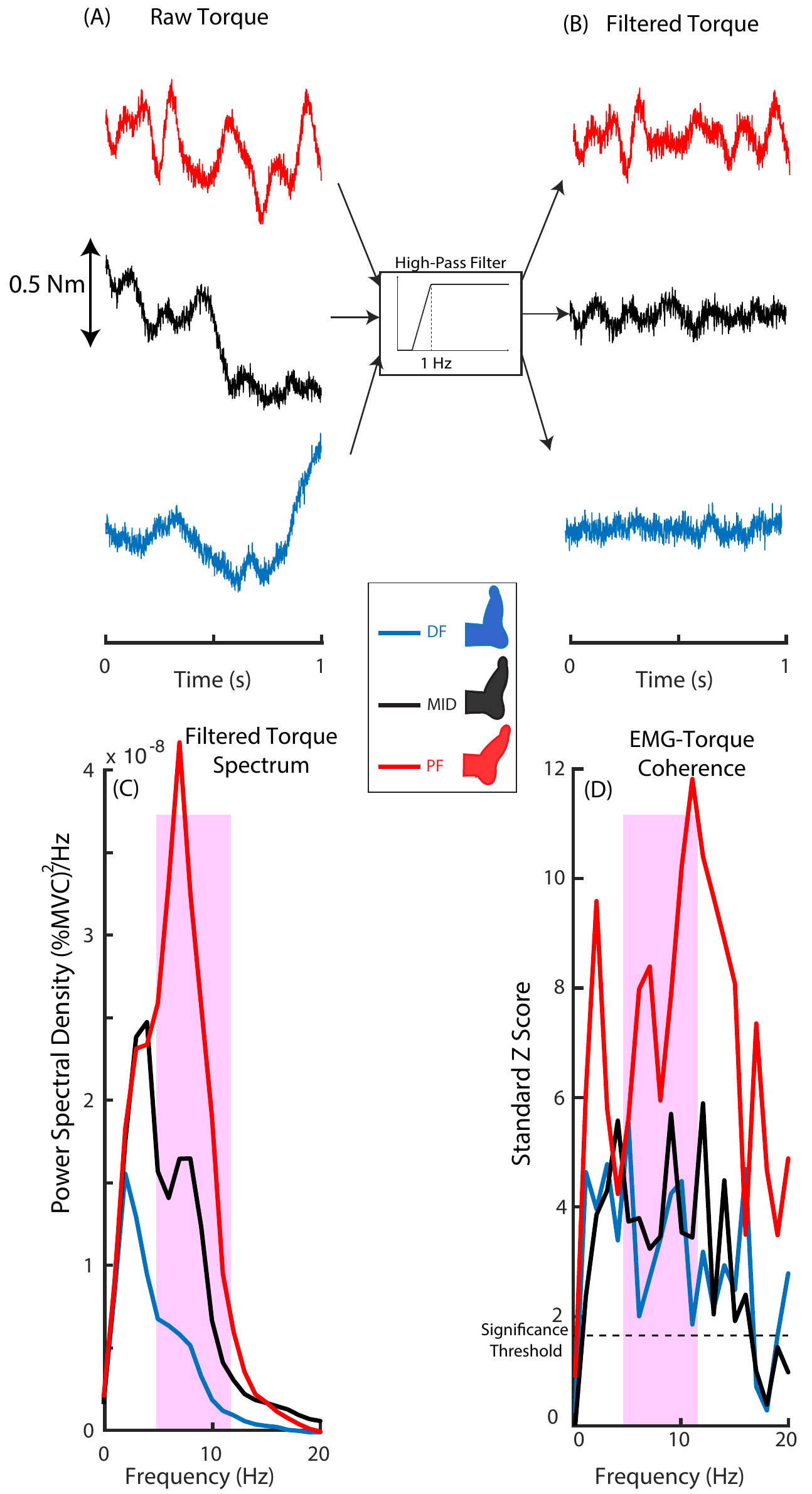}
     \caption{experimental methods: (A) Representative 1 s snapshot of torques recorded at three joint angles (torques were offset for the purpose of visual comparison); (B) torque high-pass filtered with a sixth order Butterworth filter with 5 Hz cut-off frequency; (C) power spectral density of torque calculated using a 1 s Gaussian window; (D) the EMG-torque coherence estimated using 1 s Gaussian window. The shaded area represents the physiological tremor band (5-12 Hz).}
     \label{analysisMethods}
\end{center}
\end{figure}

We also analyzed the coherence between the EMG signals of the plantarflexor muscles and torque. This helps to verify a relationship between torque tremor and tremor-band oscillations in muscle activation. First, the EMG signals were full-wave rectified, and the magnitude-squared coherence was estimated with 1 s Gaussian-tapered windows. For statistical reasons, the magnitude-squared coherence was normalized using Fisher's r-to-z transform $z_{\textrm{Fisher}}(f) = atanh(\sqrt{c(f)})$ where $c(f)$ is the magnitude-squared coherence at a given frequency $(f)$ and $atanh$ is the inverse hyperbolic tangent. Next, the transformed values were converted to standard $z_{\textrm{standard}}(f)$ scores. A bias inherent in this method was estimated (and then removed) by averaging $z_{\textrm{standard}}(f)$ over the frequency range of 100 to 300 Hz and excluding the harmonics of 60 Hz \cite{baker2001synchronization,baker2003synchronization}. Fig.~\ref{analysisMethods} illustrates these analyses on representative data.

To estimate the overall $\alpha$ drive to the calf muscles and test for possible agonist-antagonist co-activation, we analyzed the amplitude of the EMG signals of the plantarflexor and dorsiflexor muscles. We calculated the \emph{Root Mean Squared} (RMS) values of EMG signals in each trial and normalized these values to the RMS of the EMG signals recorded during the maximum contractions.

\subsection{Simulation Methods}
\subsubsection{Closed-loop Simulations of Afferented Muscle}
\begin{figure}
\begin{center}
     \includegraphics [scale=0.7]{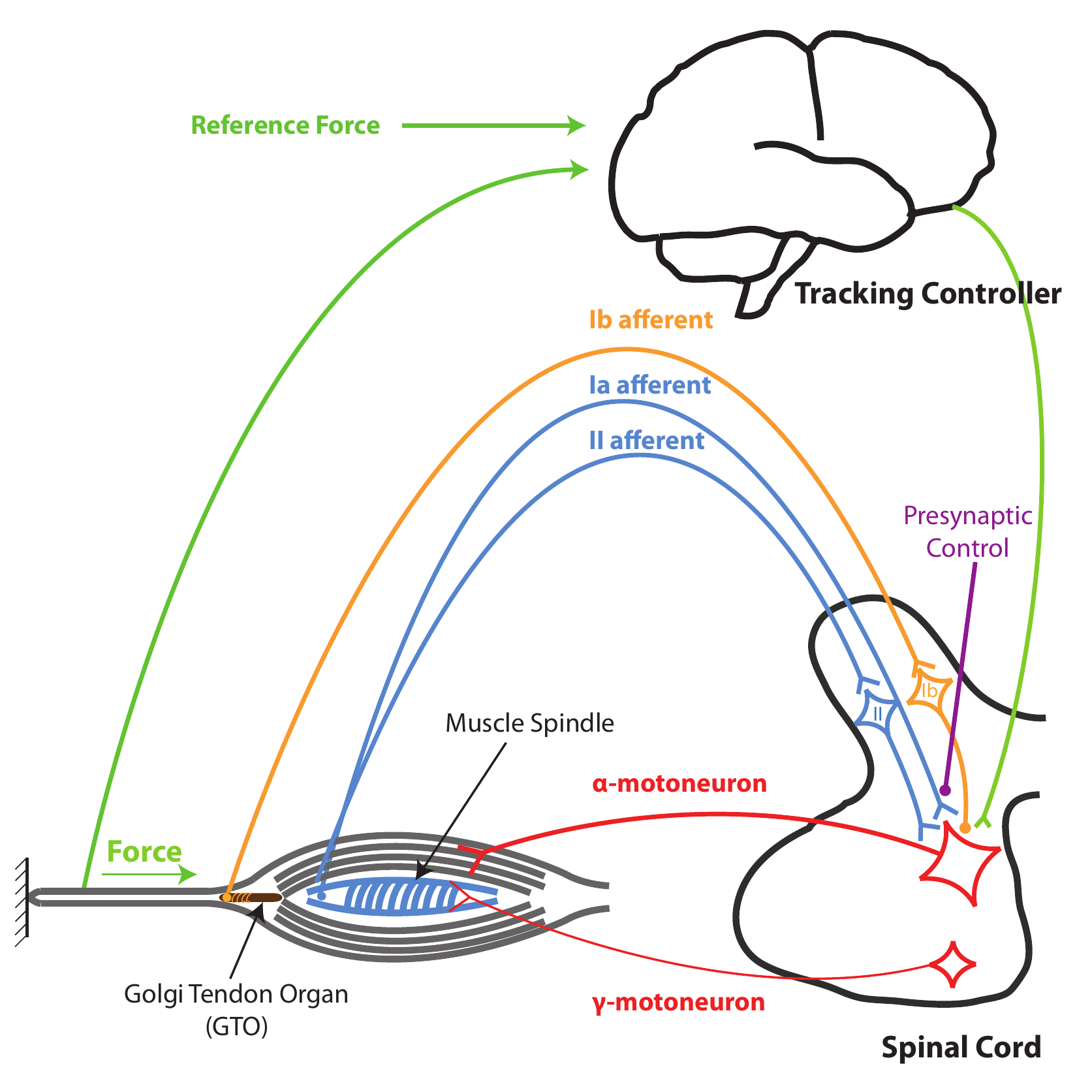}
     \caption{Schematic representation of the afferented muscle model used in simulation study. The muscle model receives neural drive from multiple sources: tracking controller, muscle spindle feedback and \emph{Golgi Tendon Organ} (GTO) feedback. Muscle spindle receive adjustable fusimotor ($\gamma$) drive and make mono- and disynaptic excitatory connections to the $\alpha$-motoneuron in the spinal cord through primary (Ia) and secondary (II) afferents, respectively. GTO makes disynaptic inhibitory connection to $\alpha$-motoneuron. Each proprioceptive feedback pathway receives presynaptic inhibition control. The tracking controller provides a control signal based on the error between the reference and output forces. The resulting muscle neural drive is passed through an activation filter that accounts for the Ca \textsuperscript{+} dynamics and produces muscle contraction. Afferent and efferent pathways have realistic time delays.}
     \label{afferentedMuscleModel}
\end{center}
\end{figure}

We used a computational model of afferented musculotendon to investigate neuromechanical factors that might contribute to modulation of physiological tremor with muscle stretch. This model was an extension of our previously published model of \citep{laine2016dynamics} by incorporating the more realistic muscle model of \citep{brown1996,cheng2000virtual,song2008computationally,song2008}, and an offset on the muscle spindle Ia firing rate to account for presynaptic inhibition mechanisms \citep{raphael2010spinal,tsianos2014useful}. Figure~\ref{afferentedMuscleModel} presents a conceptual description of this model that comprises previously published models of a musculotendon unit \citep{brown1996,cheng2000virtual,song2008computationally,song2008}, muscle spindle \citep{mileusnic2006spindle}, and \emph{Golgi Tendon Organ} (GTO) \citep{elias2014}, and a tracking controller \citep{laine2016dynamics}. Detailed descriptions of each model and model parameters can be found in the corresponding references and only brief descriptions are given here.  

The muscle model describes the known physiological properties of muscle and tendon that relate to muscle force production, including the non-linear passive behavior of muscle and tendon, force-length and force-velocity relationships of muscle, and length- and velocity-dependent activation-force dynamics \citep{brown1999measured,brown2000measured}. In this study, we specifically modeled the medial head of the gastrocnemius muscle using previously published anatomical data (Table~\ref{gastrocnemiusParams}) \citep{arnold2010,elias2014,wickiewicz1983}. We adjusted the parameters that relate to passive muscle (\textit{L\textsubscript{r1}} = 1.05) and tendon (\textit{c\textsuperscript{T}} = 100) stiffnesses to account for known changes in the intrinsic stiffness of muscle as a function of joint angle \citep{mirbagheri2000intrinsic}. We set the muscle and tendon lengths to have optimal lengths at the PF posture based on previously published passive stiffness data \citep{weiss1986position}. To simulate the experimental conditions, the muscle fiber and tendon lengths were varied as a function of joint angle based on previously published experimental measurements \citep{iwanuma2011,kawakami1998}. We obtained muscle fiber length changes of 0.6 and 1.0 cm, and tendon length changes of 0.2 and 0.4 cm from PF to MID and from PF to DF joint angles, respectively.

The feedback control system that generates neural drive to the muscle (see Fig ~\ref{afferentedMuscleModel}) consisted of a tracking controller and proprioceptive feedback pathways. The tracking controller generates a command signal based on the error between the actual force output and reference force \citep{laine2016dynamics}. This controller was not designed to represent specific neural pathways, but it ensures successful force tracking \citep{laine2016dynamics}. Proprioceptive feedback pathways consisted of Ia and II afferent feedback arising from muscle spindles and Ib afferent feedback from GTO. The contribution of each proprioceptive feedback pathway was controlled by ``presynaptic inhibition control'' input \citep{raphael2010spinal,tsianos2014useful}. Each feedback pathway was appropriately delayed based on the simulated distance of 0.8 m between the spinal cord and muscle as well as the conduction velocities of the respective afferent and efferent fibers \citep{elias2014} (Table ~\ref{gastrocnemiusParams}). \emph{Signal Dependent Noise} (SDN) was added to the  neural drive such that the force variability of the model matched approximately that recorded experimentally (coefficient of variation of $\approx$3 \%). The SDN was modeled as zero-mean signal with variance of 0.3 generated by low-pass filtering white noise with a 4th-order Butterworth filter at 100 Hz. The neural drive was first passed through an activation filter, which accounts for low-pass dynamics of calcium uptake in skeletal muscle \citep{song2008}. The resulting activation signal induces muscle contraction and associated dynamics within the musculotendon unit.

\begin{table}
\caption{Model parameters of the medial head of the gastrocnemius muscle}
\label{gastrocnemiusParams}
\centering
\begin{adjustbox}{max width=\textwidth}
\def\arraystretch{2}
\begin{tabular}{c c}
\hline 
\ Mass (g) & 75 \\[1ex]

\ Optimal fiber length (cm) & 10.1 \\[1ex] 

\ Tendon length (cm) & 23.5  \\[1ex] 

\ Pennation angle (deg) & 5 \\[1ex] 

\ Ia afferent conduction velocity (m/s) & 64.5 \\[1ex] 

\ II afferent conduction velocity (m/s) & 32.5 \\[1ex] 

\ Ib afferent conduction velocity (m/s) & 59 \\[1ex] 

\ $\alpha$-motoneuron conduction velocity (m/s) & 56\\[1ex]

\ Synaptic delay (ms) & 2\\[1ex]
\hline
\end{tabular}
\end{adjustbox}

\end{table}

\subsubsection{Simulation Protocols}
We simulated MVC and passive force trials at each joint angle similar to those acquired experimentally. To do so, we used a ramp-and-hold reference \emph{force} input consisting of a 1 s zero activation, 1 s ramp-up, and 13 s maximum activation phases. We removed all the feedback components (proprioceptive feedback and tracking controller). We calculated the passive and MVC forces as the average force during the last 0.5 s of the zero activation and maximum activation phases, respectively.

\begin{figure}
\begin{center}
     \includegraphics [scale=0.2]{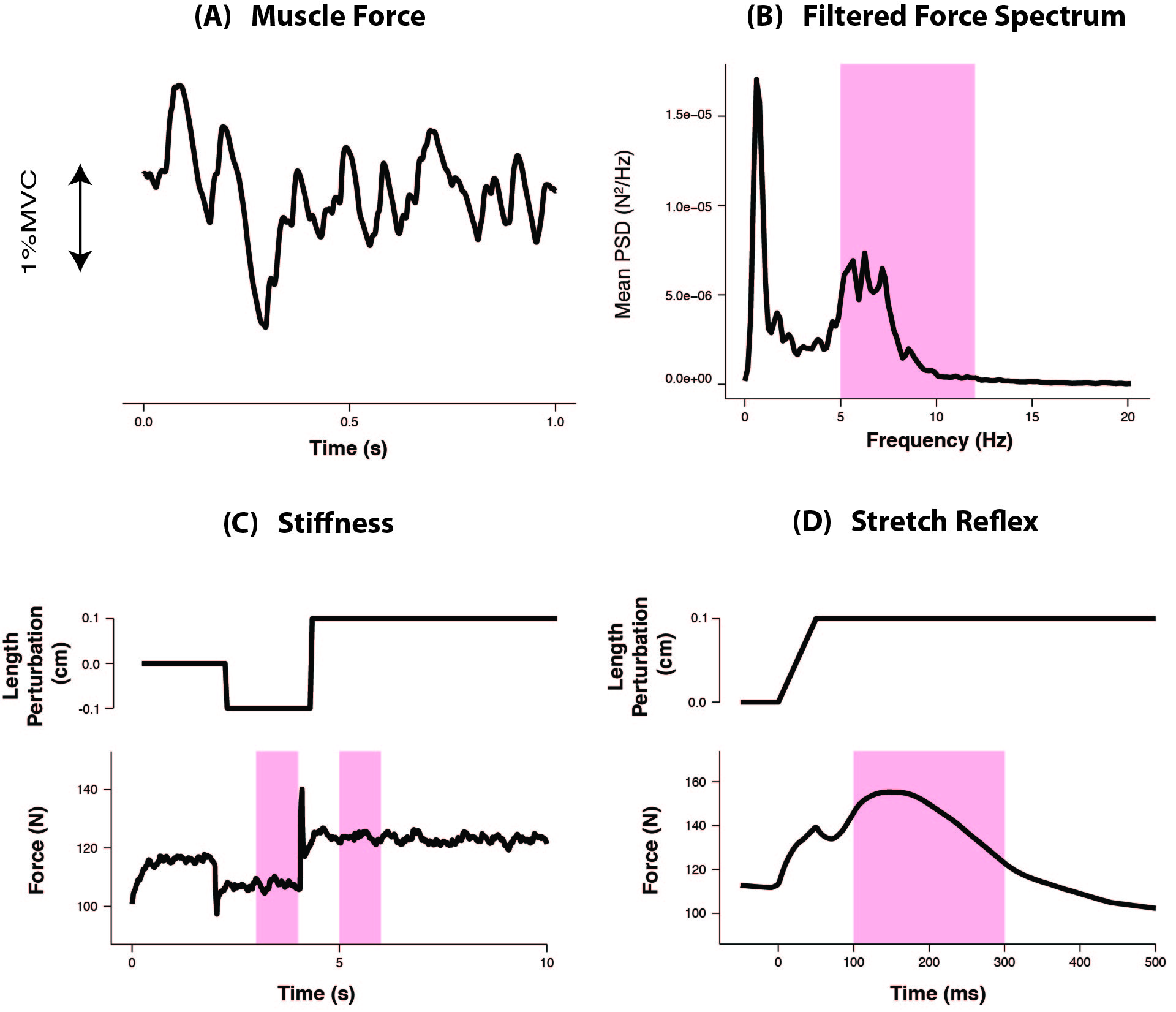}
     \caption{Simulation methods: (A) Representative 1 s snapshot of the simulated muscle force; (B) representative power spectral density of muscle force. The shaded area represents physiological tremor range; (C) simulated muscle stiffness experiment. The difference in muscle force between the shortened and lengthened phases (shaded areas) was obtained to compute muscle stiffness (see Methods); (D) simulated stretch reflex experiment. Reflex amplitude was obtained as the peak force value in the 100-300 ms window after the perturbation onset (shaded area). }
     \label{methodsSimulation}
\end{center}
\end{figure}

Next, we simulated tonic, isometric contraction trials for 100 s at 30\% MVC (Figure \ref{methodsSimulation}). We analyzed the last 90 s of the simulated force data to exclude the transient response. The simulated force data were analyzed in the same way as the experimental data.

To explore contributions of mechanical and neural factors to modulation of physiological tremor as a function of joint angle, we performed a sensitivity analysis. First, we performed a Monte-Carlo analysis of 20 trials to obtain the statistics of the estimates by simulating tonic, isometric contractions. Second, given the experimental reports that Ia afferent feedback is important for the modulation of physiological tremor \citep{lippold1971physiological}, we manipulated the gain of the Ia afferent feedback as a function of joint angle, either through presynaptic control of Ia afferent or fusimotor drive to the muscle spindle. We increased presynaptic inhibition by 20\% (i.e., decreased presynaptic control input levels by 20\%) at the DF angle compared to the other two joint angles to mimic the changes in H-reflex amplitudes observed experimentally \citep{hwang2002assessment}. We also studied the effect of fusimotor drive by independently varying $\gamma$-dynamic and $\gamma$-static drive from 10 to 250 pps with increment of 20 pps at each joint angle. In this paper, we present the ``optimal'' set of $\gamma$-dynamic and $\gamma$-static drives that produced behavior most similar to that observed experimentally.
\vspace{5mm}

\noindent
\textit{Quantification of Muscle Stiffness}

\noindent
We quantified changes in the mechanical stiffness of muscle as a function of joint angle to explore potential mechanical explanations for modulation of physiological tremor. The afferented muscle model was first deafferented by removing all feedback components. Next, muscle length perturbations of four amplitudes (0.05, 0.1, 0.15, and 0.2 cm) were applied directly to the musculotendon unit during sustained contraction at 30\% MVC, each 20 times. For each perturbation, the musculotendon unit was shortened for 2 s and then lengthened to the original length over 50 ms (Figure ~\ref{methodsSimulation} C top panel). The difference in the mean muscle force (F\textsubscript{diff}) before and after the perturbation was calculated in a time-window of 1 s (shown in pink shaded areas in Figure ~\ref{methodsSimulation} C bottom panel).

Stiffness at each joint angle was quantified as the slope of the first-order regression line fitted between perturbation amplitudes and  $F_{\mbox{diff}}$ (on 20$\times$ 4 = 80 points). We used first-order regression based on the HC4 method \citep{wilcox2012}, which allows for heteroscedasticity (unequal variance among groups) and therefore provides an accurate estimate of the confidence interval of the slope (i.e. stiffness). 
\vspace{5mm}

\noindent
\textit{Quantification of Stretch Reflex}

\noindent
We simulated a stretch reflex experiment to demonstrate the consistency of the model with previous experimental observations. Here, 20 direct ramp-and-hold perturbations (amplitude of 0.1 cm and velocity of 2 cm/s, shown in Figure ~\ref{methodsSimulation} D top panel) were applied to the afferented muscle model at each simulated muscle length. The reflex amplitude (phasic response) was calculated as the difference between the mean force level  in a 50-ms window before the perturbation and peak muscle force within a 100 - 300 ms window after the perturbation onset (shown in pink shaded areas in Figure ~\ref{methodsSimulation} D bottom panel). 

\subsection{Statistical Analyses}
Statistical analyses were performed to identify the dependence of outcome variables (e.g., total power, physiological tremor power, and stretch reflex amplitude) on joint angle (independent variable). For experimental data, we used the paired, Wilcoxon signed rank test (built-in function in R) on the normalized data. 

For simulation results, we used an unpaired version of a heteroscedastic one-way ANOVA for means \citep{wilcox2012}. If a significant interaction was found between an outcome variable and joint angles, corresponding post-hoc analysis that accounts for multiple comparisons was performed to identify significant differences between pairs (Welch's method, lincon function in "WRS2" package) \citep{wilcox2012}. 

We performed the statistical analyses in the R environment for statistical computing (The R Foundation for Statistical Computing, Vienna, Austria). The significance level was set to p $<$ 0.05.  

%% file: results.tex
\section{Results}


\begin{figure}
\begin{center}
     \includegraphics [scale=1.5]{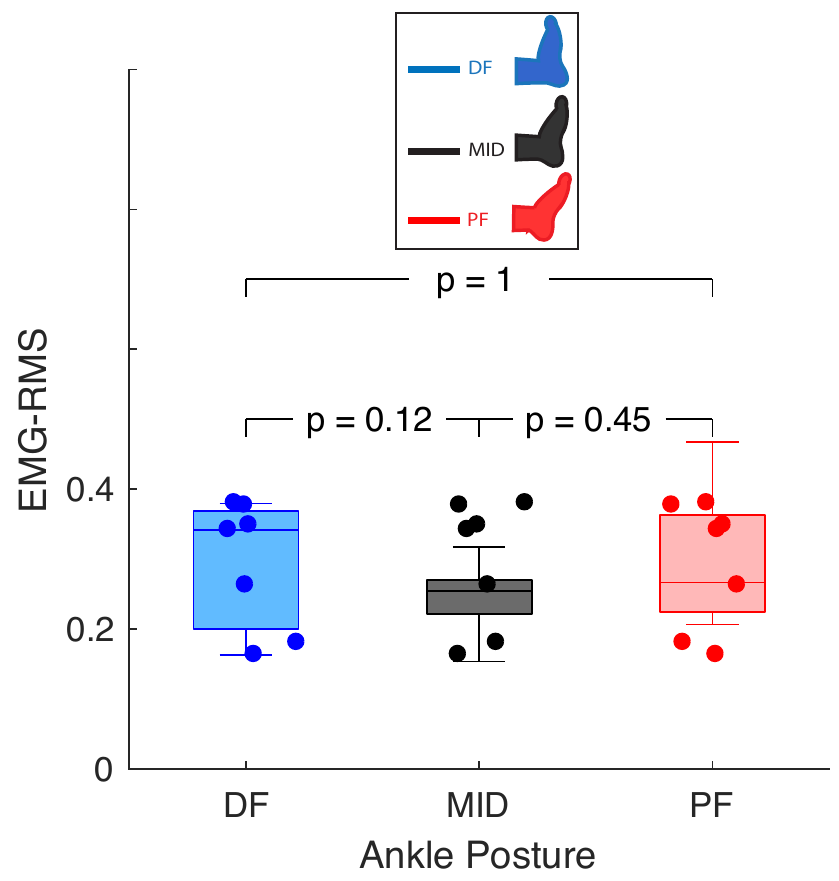}
     \caption{RMS level of the EMG signals of the calf muscles (average of gastrocnemius medialis, gastrocnemius lateralis, and soleus) does not change significantly as a function of joint posture. Each point represents the EMG RMS level of a subject.}
     \label{alphaDrive}
\end{center}
\end{figure}

\begin{figure}
\begin{center}
     \includegraphics [scale=0.8]{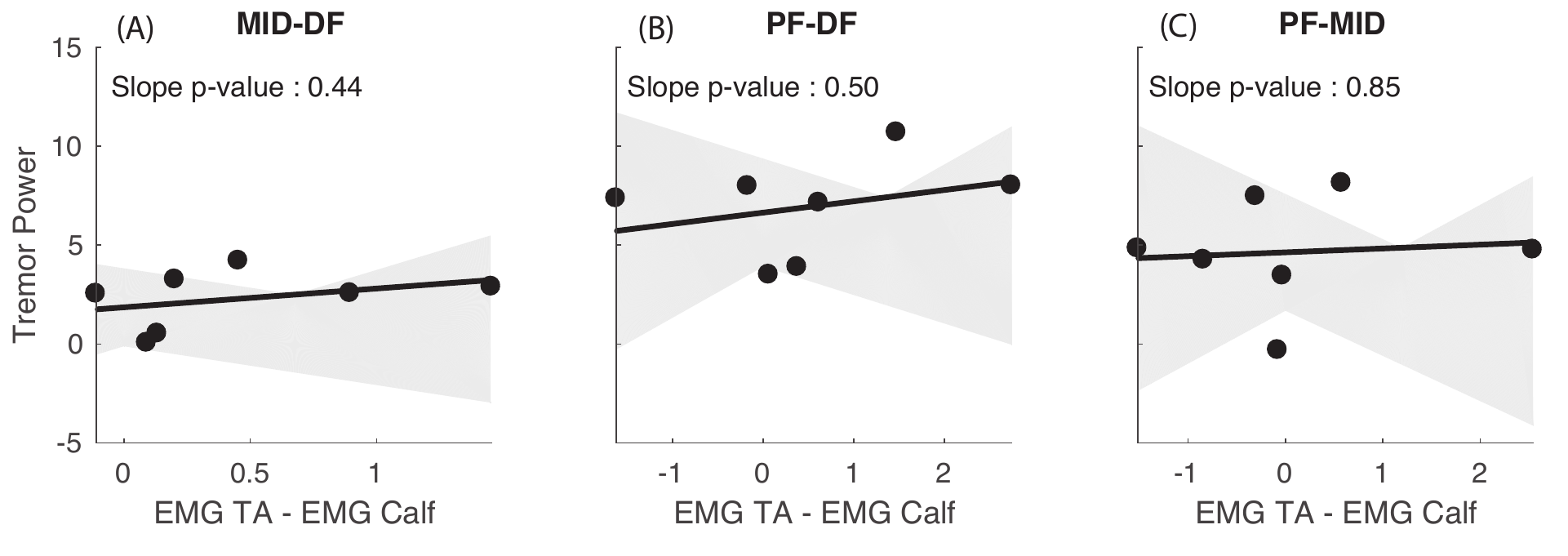}
     \caption{Changes in physiological tremor power were not correlated with possible co-activation of the tibialis anterior and calf muscles: Panels show changes in tremor power as a function of changes in the difference of the tibialis anterior and calf muscles RMS EMG values: (A) between the MID and DF postures; (B) between the PF and DF postures; (C) between the PF and MID DF postures.}
     \label{co-activation}
\end{center}
\end{figure}

\begin{figure}
\begin{center}
     \includegraphics [scale=1.5]{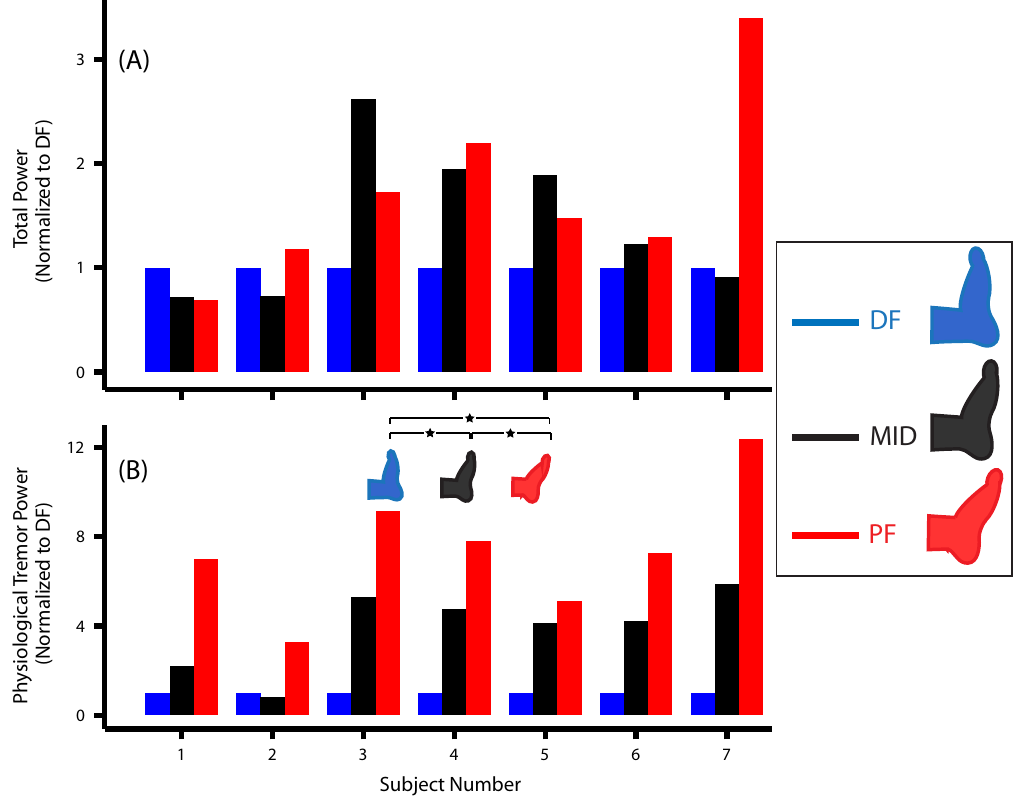}
     \caption{Torque power as a function of joint angle for individual subjects. (A) torque total power shows no consistent pattern of change as a function of joint angle; (B) torque power in the physiological tremor range (5-12 Hz) increases from dorsiflexed to plantarflexed angles. }
     \label{totalTremorPower}
\end{center}
\end{figure}

To investigate potential changes in the level of neural drive to muscle as a function of joint angle, we calculated the average RMS EMG values across the calf muscles, i.e. the three ankle plantarflexor muscles. We found no significant difference in the RMS values across the three joint angles ($p >=$ 0.12, Figure~\ref{alphaDrive}). Next, we computed the RMS values of the tibialis anterior muscle to examine potential changes in the level of co-contraction as a function of joint angle. Once again, we found no differences in the level of muscle activation across joint angles ($p >=$ 0.12). We also analyzed the individual calf muscles using the same technique and we found no significant differences in the level of muscle activation across joint angles ($p >=$ 0.12). These results show that there was no angle-dependent modulation of plantarflexor and dorsiflexor muscle activation.

Next, we tested for potential contribution of co-activation of the ankle dorsiflexor in modulation of physiological tremor. To this end, we normalized the EMG RMS values to those of the DF posture and computed the difference between the calf (average across the three plantarflexors) and tibialis anterior muscle RMS EMG as a function of ankle angle. We correlated changes in physiological tremor with the difference in the EMG RMS values. Figure~\ref{co-activation} demonstrates that this correlation was not significant. In fact, the slope of the fit line was not significantly different than zero ($p>0.44$). This result reinforces the lack of angle dependent changes in tibialis anterior/calf muscles coactivation.

To ensure that physiological tremor was related to the plantarflexor's activities, we computed the coherence between the plantarflexor EMGs and torque. The EMGs showed significant coherence (larger than significance threshold shown by the dotted line) with plantarflexion torque in the physiological tremor range in all subjects (Figure~\ref{analysisMethods} (D) shows a typical coherence plot). This validates the use of the plantarflexor muscle in our simulations for investigating physiological tremor. Since on average, the medial head of gastrocnemius muscle showed higher coherence compared to the other two muscles, we opted to use the architectural parameters of the gastrocnemius medialis in the simulation of the afferented muscle model.

\subsection{Physiological Tremor Increased with Muscle Shortening.}
Figure~\ref{totalTremorPower}(A) demonstrates that there was no consistent pattern of change in total torque power as a function of joint angle (p $=$ 1, 0.125, and 1, for MID-DF, PF-DF, and PF-MID, respectively). However, the amplitude of the physiological tremor (5-12 Hz) showed a consistent pattern across all subjects in Figure~\ref{totalTremorPower} (B). Specifically, physiological tremor was larger in the PF than MID, and DF postures ($p =$ 0.031 and $p =$ 0.016, respectively) and in the MID than DF posture ($p =$ 0.016). These results demonstrate that modulation of torque fluctuations was frequency-specific, and that power in the physiological tremor range increased as the muscle shortened.

\subsection{Simulations of Afferented Muscle to Evaluate Potential Neuromechanical Factors to Explain Experimental Results}
To examine potential origins of physiological tremor modulation with respect to joint angle, we simulated tonic, isometric contraction trials at the muscle and tendon lengths corresponding to the three ankle angles tested experimentally (Figure \ref{methodsSimulation}(A)). Simulations of afferented muscle can replicate physiological tremor, similar to the experimental findings (Figure \ref{methodsSimulation}(B)).

We used a systematic process of elimination to identify the mechanical and neural factors responsible for the angle dependent modulation of physiological tremor (Figure~\ref{simulationResults}(A-1)). The considered factors included musculotendon mechanics, presynaptic gain of Ia afferent feedback, and fusimotor drives. 

\subsubsection{Effect of Mechanical Factors}

Figure~\ref{simulationResults}(A-2) shows the changes in physiological tremor power as a function of joint angle when all parameters related to afferent feedback were kept constant across muscle lengths. Contrary to the experimental results, we found that mechanical factors by themselves resulted in a decrease in physiological tremor when the muscle shortened. This indicates that our experimental observations cannot be explained by neuromechanical factors related to changes in muscle length, such as mechanical properties of the muscle-tendon system and changes in the background activity of afferents. 

\subsubsection{Effect of Presynaptic Inhibition}
To explore effects of afferent feedback modulation on physiological tremor amplitude, we ran simulations with different afferent feedback gains. A previous study demonstrated that the level of presynaptic inhibition of Ia afferents can vary with joint angle \citep{hwang2002assessment}. We tested this by increasing the presynaptic inhibition by 20\% at DF (from presynaptic control level of -0.1 to -0.12) compared to the other two joint angles \citep{hwang2002assessment}. This modulation had little impact on physiological tremor amplitude (Figure~\ref{simulationResults} (C)), suggesting that modulation of presynaptic inhibition of Ia afferent feedback does not suffice to reproduce our experimental results.

\subsubsection{Effect of Fusimotor Drive}
Next, we tested the influence of fusimotor drive modulation. We first increased $\gamma$-dynamic as the muscle was shortened. As shown in Figure~\ref{simulationResults}(A-4), this was not sufficient to replicate experimental observations either. From all the tested combinations of $\gamma$-dynamic and $\gamma$-static (see methods), we found that only increases in $\gamma$-static resulted in greater physiological tremor in the plantarflexed posture. Figure~\ref{simulationResults}(A-5) demonstrates that increasing $\gamma$-static from 10 pps to 70 pps and to 190 pps in the DF, MID and PF angles with a fixed $\gamma$-dynamic of 110 pps results in a consistent monotonic increase in physiological tremor amplitude similar to the pattern we identified experimentally. Further, physiological tremor amplitude was significantly larger at the PF angle compared to the other two angles (p $<$ 0.01 for both comparisons) and in the MID compared to the DF angle (p $<$ 0.01). These results suggest that the modulation of physiological tremor observed experimentally could stem from modulation of $\gamma$-static fusimotor drive with muscle/tendon stretch. 

\subsubsection{Model Validation: Reproducing Muscle Stiffness}
To confirm the validity of the chosen model parameters, we tested our model to see if it exhibited the known angle dependent changes in muscle stiffness \citep{mirbagheri2000intrinsic}. Figure~\ref{simulationResults}(B-2) shows that the model's muscle stiffness increased from the PF to MID and to DF. This is consistent with a previous experimental report that showed that ankle stiffness increased monotonically from $27 ^{\circ}$ plantarflexion toward full dorsiflexion as shown in Figure~\ref{simulationResults}(B-1) \citep{mirbagheri2000intrinsic}. This further supports our conclusion that modulation of physiological tremor as a function of joint angle cannot be explained by alterations in mechanical properties of the musculotendon unit. 

\subsubsection{Model Validation: Reproducing Stretch Reflex Response}
We further tested if the known effect of joint angles on stretch reflex amplitude \citep{mirbagheri2000intrinsic} emerges from our simulation when fusimotor parameters are set to replicate the experimental behavior of physiological tremor. Figure~\ref{simulationResults}(C-2 and C-5) shows that the stretch reflex response increased from PF to MID and to DF, which is again consistent with previously reported observations Figure~\ref{simulationResults}(C-1) \citep{mirbagheri2000intrinsic}. This result confirms that our model produces physiologically realistic modulation of \underline{both} physiological tremor and stretch reflex amplitude.

\subsubsection{Experimental Study: Validation of Stiffness and Stretch Reflex Response}
We recruited one subject to further validate, simultaneously, the angle-dependence of both stretch reflex amplitudes and muscle stiffness. We placed the ankle at the PF, MID and DF postures and applied pulse perturbations to the ankle. The peak amplitude of the pulses was $1.15 ^{\circ}$. The duration of each pulse was 1s and time interval the between pulses was a random uniform variable with minimum of 2s and maximum 4s.

\begin{figure}
\begin{center}
     \includegraphics [scale=0.7]{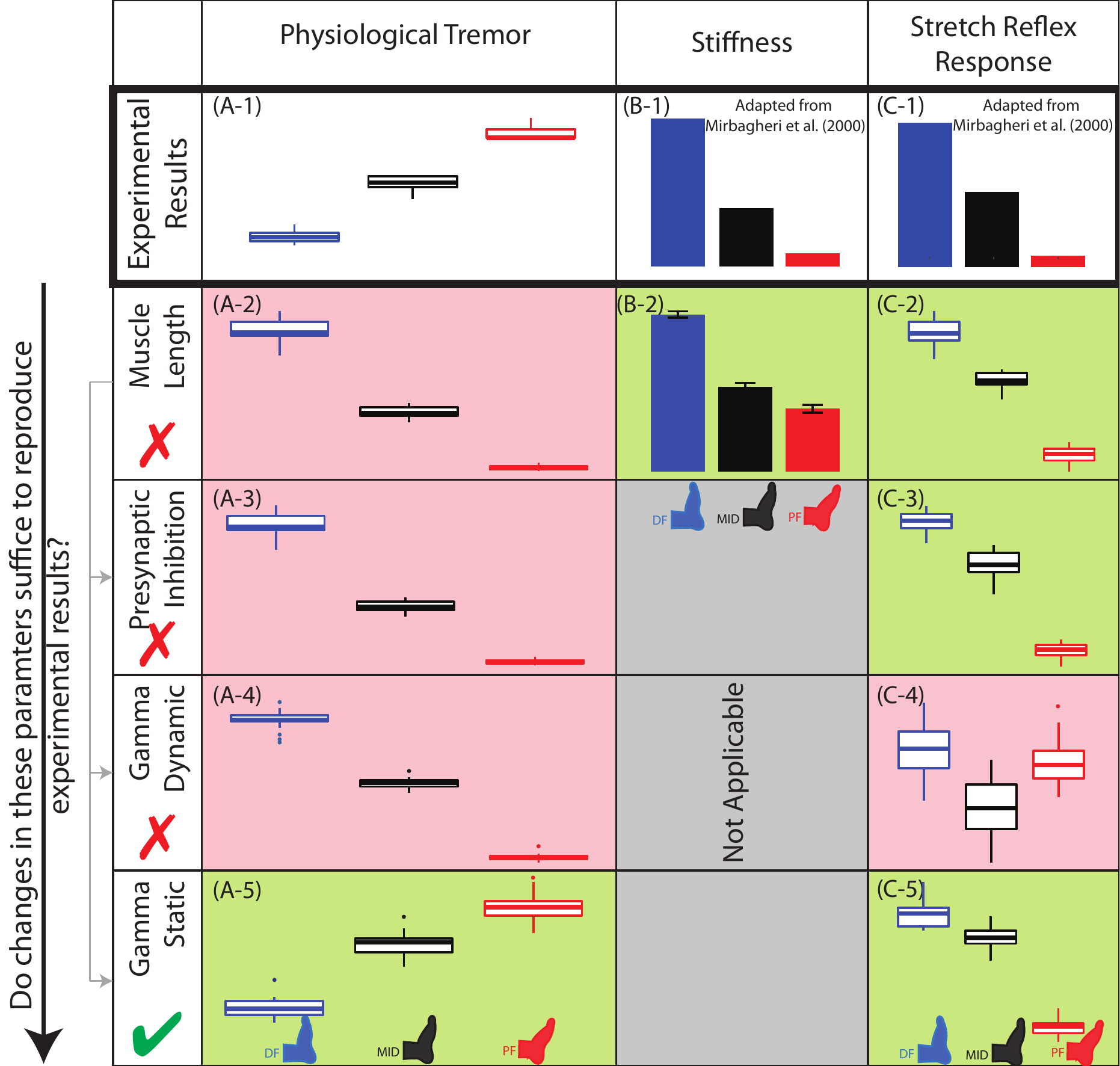}
     \caption{Simulation results showing that only an increase in $\gamma$-static is sufficient to replicate the experimentally observed muscle length dependence of physiological tremor. We used an afferented muscle model and systematically tested different factors that can contribute to physiological tremor.  By process of elimination we identified the sufficient conditions under which the model replicates experimental observations. The first column evaluates ankle-angle dependent changes in physiological tremor in healthy adults(A-1); and in simulation when (A-2) only muscle length was varied and the neural gains were held constant, when (A-3) presynaptic inhibition was increased at the dorsiflexed posture, when (A-4) $\gamma$-dynamic was increased at the plantarflexed posture, and when (A-5) $\gamma$-static was increased at the plantarflexed posture. The second column shows changes in muscle stiffness across joint angles (i.e, muscle lengths) in healthy adults (B-1) and in our model (B-2). The third column shows changes in stretch reflex responses in healthy adults(C-1) as well under the simulated conditions described above (C-2 to C-5).}
     \label{simulationResults}
\end{center}
\end{figure}

We used the technique for the simulation data to identify stiffness and reflex responses. Stiffness was 1.88 $\pm$ 0.24 (Nm/${\circ}$) at the DF, 1.44 $\pm$ 0.16 (Nm/${\circ}$) at the MID, and 0.97 $\pm$ 0.16 (Nm/${\circ}$) at the PF postures. Stretch reflex response was 0.94 $\pm$ 0.60 (Nm) at the DF, 0.37 $\pm$ 0.20 (Nm) at the MID, and 0.15 $\pm$ 0.05 (Nm) at the PF postures. The direction and magnitude of changes in stiffness and stretch reflex response are very comparable to those of the simulation model and the previous works (see Figure~\ref{simulationResults}).

%% file: discussion.tex
\section{Discussion}

We studied the dependence of physiological tremor on the length of the muscle-tendon complex. We hypothesized that tremor amplitude would be proportional to the gain of the stretch reflex loop and therefore increase when the muscle was lengthened. Surprisingly, we found the opposite: tremor amplitude increased when muscle was shortened. We explored potential mechanical and neural mechanisms underlying this observation by simulating a model of an afferented muscle-tendon system controlled in closed-loop. Using this simulation, we were able to exclude basic mechanical properties of muscle responsible for our experimental observations. Importantly, we found that replicating experimental results required increasing the activation of $\gamma$-static fusimotor drive as the muscle was shortened. This previously-undescribed relationship between fusimotor drive and muscle length (independent of movement, changes in $\alpha$-drive, or antagonist muscle activity) may begin to fill important gaps in our knowledge of fusimotor activity during voluntary actions. It may also provide novel avenues for experimental characterization of fusimotor activity, or its dysfunction in the context of injury or disease.

\subsection{Limitations of the Study}
As with any computational model, our results help to understand experimental observations, but are not direct mechanistic proofs. No simulation is complete \citep{valero2009computational}, and we intentionally used a simple implementation which lacks biophysical models of spiking neurons, Renshaw cells, or supraspinal circuitry. Our controller, though reminiscent of a cortical long latency reflex loop in conception, is not a neurophysiological model of descending control, and we only consider a single isolated muscle rather than a coordinated group of agonists/antagonists, as is almost always the case in reality.

The surface EMG measurement suffers from several issues including: cross-talk especially in ankle muscles \cite{winter1994crosstalk}, dispersion of action potentials \cite{phinyomark2012usefulness}, changes in muscle geometry, etc. We should emphasize that while we used surface EMG measurement in this study, the precise correspondence between EMG and torque was not the primary focus. Rather, we used EMG to demonstrate that $(i)$ tremor had neural origin (versus mechanical) by showing the torque variability in the tremor frequency range is coherent with EMG, $(ii)$ the average activation level in calf muscles is held constant as a function of joint angle, and $(iii)$ antagonistic co-activation was minimal and did not have a posture-dependent component. While we expect these issues to have small effects in our final conclusions, it is within interest to use intramuscular EMG recording to fully characterize the EMG-torque coherence.

Even with these limitations, our prediction of neural (rather than purely mechanical) responses to altered muscle length remains a valid and particularly a straightforward explanation for our experimental observations. In this sense, the simplicity of our simulation is a strength, since basic pathway gain adjustments can explain many different aspects of tremor modulation.

In future, it is within our goals to include more anatomically and physiologically faithful structures of the tendon-driven limbs. More specifically, we are interested to include multiple synergistic and antagonistic muscles actuating the joint. We will use muscle models scaled to the three calf, and tibialis anterior muscles. We will include more spinally mediated neural pathways connecting these muscles. This includes classical autogenic and heterogenic neural (heteronomous spindle pathways shared between the calf muscles and reciprocal inhibition between the calf and tibialis anterior muscles) \cite{nichols2016neural}, Renshaw cells, and propriospinal interneurons \cite{raphael2010spinal}.
Inclusion of $\beta$ motoneurons may also expand our understanding of neural control of mechanoreceptors. These neurons innervate both intrafusal and extrafusal muscle fibers \cite{emonet1992comparison,jami1982quantitative,manuel2011alpha,burke1977histochemical}. However, our current understanding of these neurons is very limited and further research into this matter is required to pave the way.

Moreover, as the next logical step, we will study physiological tremor across different muscle activation and co-activation levels during tonic and time-varying contraction profiles. Based on the current knowledge of neuromuscular control, we can hypothesize that physiological tremor increases with muscle activation due to $\alpha-\gamma$ co-activation and facilitation of recruitment of larger motoneurons. We also expect during co-activation of the ankle plantarflexor/dorsiflexor muscles, reciprocal inhibition plays an important role in modulating physiological tremor as a function of joint angle. Thus, when dorsiflexor is stretched (PF angle) or more activated, it has a larger spindle afferent activity which in turn inhibits the motoneurons of the calf muscles. The degree to which this inhibition can alter the physiological tremor is subject to future study and validation against experimental data. This will provide critical insight into the coordination of muscles in force production tasks.

\subsection{Muscle Length Dependence Modulation of Fusimotor Drive}

Direct recording of fusimotoneurons in humans is not yet possible. Nevertheless, recording from afferent nerves using microneurography has indirectly revealed that many factors can influence fusimotor activity in humans, independently of the physical state of the muscle \citep{dimitriou2016enhanced,ribot2000increased,vallbo1981independence,prochazka1985fusimotor}.

Although a direct dependence of fusimotor drive on muscle length alone has not been reported previously, it has been shown that activity of fusimotoneurons (specifically the $\gamma$-static motorneurons) co-varies with the ankle joint angle during locomotion in decerebrate cat \citep{taylor2000distinctive,taylor2006static,ellaway2015muscle}. The favored interpretation of this is the well-known hypothesis of simultaneous $\alpha-\gamma$ co-activation, \cite{loeb1984control}. Our study suggests that the length of the muscle itself (independently of changes in $\alpha$-drive) may contribute significantly to the decoupled modulation of fusimotor activity (and more specifically the static fusimotor activity).

Our findings of independent modulation of $\gamma$-static with muscle length would also explain the common interpretation that $\alpha-\gamma$ co-activation as a function of joint angle is a means to compensate for shortening of intrafusal fibers. This has been demonstrated by (1) \citep{lan2012fusimotor} that have shown the necessity of $\gamma$-static control during arm locomotion with controlled spindle sensitivity using a simulation study ; and (2) \citep{grandjean2014model,grandjean2017emergence} who have shown a similar $\gamma$-static mechanism during wrist movement. Importantly, the results of these modeling studies have almost exclusively been interpreted in the context of $\alpha-\gamma$ co-activation.  

Deviations from strict $\alpha-\gamma$ co-activation have already been reported, e.g. in \cite{dimitriou2016enhanced} during active naturalistic movements of the wrist joint and in \citep{ribot2009fusimotor} during passive movement of the ankle joint. More recently, it has been demonstrated that adjustment of spindle sensitivity via motorneuron efferent control is not only proportional to the $\alpha$ drive of the parent muscle but also inversely proportional to the α drive of the antagonistic muscle thanks to the reciprocal inhibition pathways. Thus, a more general form of the $\alpha-\gamma$ co-activation reflects the balance of activity between antagonistic muscles \cite{dimitriou2014human}. Nevertheless, our findings of length-dependent, differential modulation of fusimotor drive (static only) that does not comply with $\alpha-\gamma$ co-activation but provides a mechanistic approach to explain some of such deviations.

\subsection{Alternative Mechanisms}

Previous experimental studies have revealed that ankle joint stiffness, on average, decreases with plantarflexion until $27 ^{\circ}$ plantarflexion \citep{stiffnessLPV2013,jalaleddini2015parametric}. Thus, a reduction of muscle stiffness with muscle shortening during plantarflexion might explain our results \citep{davidson2016fundamental}. 

Our results come from the systematic exploration of the capabilities of our model under different conditions. For example, if we only change muscle length (Figure~\ref{simulationResults}(row 2)) we see its consequence to physiological tremor, intrinsic muscle stiffness and stretch reflex response. This case demonstrates that our model is capable of replicating the intrinsic stiffness of muscles in this task (B-2), yet did not replicate the observed physiological tremor modulation when $\gamma$-static was constant (A-2). 

Rows 3 and 4 in Figure~\ref{simulationResults} demonstrate that how changing the gain of the afferent feedback (while also including changes in muscle length) did not replicate experimental observations of physiological tremor either (A-3, presynaptic inhibition, and A-4, $\gamma$-dynamic).

Consequently, the mechanical properties (muscle stiffness), presynaptic inhibition, and $\gamma$-dynamic fusimotor drive ---by themselves--- do not explain all experimental observations of physiological tremor, muscle stiffness and stretch reflex response. These results strengthen our conclusion that fusimotor drive decoupled from $\alpha$ drive (i.e., changes in $\gamma$-static with changes in muscle length) is a potentially important mechanism.

\subsection{Stretch Reflex and Physiological Tremor Amplitudes Can Be Decoupled}
Physiological tremor is characterized by synchronized firing of motoneurons in the range of 6-12 Hz, which is believed to emerge from self-sustaining oscillations of excitation in the stretch reflex loop \citep{hagbarth1979participation,lippold1970oscillation,mcauley2000physiological,takanokura2005neuromuscular}. Previous studies have shown that the amplitude of physiological tremor increased whenever afferent gain increased, for example by cognitive/perceptual aspects of force control tasks \citep{laine2015motor}, or by direct manipulation of afferent nerves \citep{christakos2006parallel}.

Accordingly, it is not unreasonable to expect that physiological tremor amplitude should increase when muscle was lengthened, as muscle lengthening is known to increase stretch reflex amplitudes \citep{mirbagheri2000intrinsic}---yet this relationship has not been conclusively established as obligatory. In contrast to this line of reasoning, our data demonstrate decoupling of physiological tremor and stretch reflex amplitudes.

Decoupling of measures of afferent sensitivity has been previously reported \citep{maluf2007reflex}. We have even recently documented that \emph{increases} in $\gamma$-static fusimotor drive can actually \emph{decrease} the stretch reflex amplitude \citep{jalaleddiniNeuromorphic2017}. Our results here now provide further evidence to suggest that the amplitudes of the stretch reflex and physiological tremor are separable---likely because they are driven by distinct, but complementary and sometimes overlapping, afferent information. This is likely because stretch reflex amplitude depends on sensitivity/gain along different pathways, each of which operates on a slightly different time-scale, and responds to different stimuli (length, velocity, force, etc). 

We propose that assessment of muscle force variability might be an important avenue for future investigation of reflex modulation in health and disease. While we have identified a potential fusimotor strategy that suffices to replicate changes in physiological tremor, muscle stiffness, and stretch reflex response, it will require further research to fully characterize the relationship between fusimotor drive and physiological tremor.

\subsection{Clinical Implications}
Rhythmic oscillation in execution of motor tasks is clinically significant in a variety of disorders including Parkinson's disease \citep{ko2015force,vaillancourt2001regularity}, Dystonia \citep{xia2007modulation,chu2009force}, bruxism \citep{laine2015jaw}, essential tremor \citep{heroux2010effect,gallego2017neural}, etc. Moreover, abnormal gains and thresholds in the spinal loop circuitry have been documented in pathologies such as spinal cord injury, stroke, and multiple sclerosis  \citep{o1998abnormal,jobin2000regulation,kamper2000quantitative,tuzson2003spastic}.

The results of this study suggest that the variability in isometric ankle torque production at different dorsi- plantarflexion angles could serve as a simple, non-invasive, quick and clinically-practical tool to characterize the integrity of afferent and reflex circuitry. Thus, it can serve as a means to explore neuromechanical coupling and reflex pathway gains during natural behavior---and without the need for mechanical perturbation or nerve stimulation. Because the amplitude of physiological tremor is negligible for most motor control tasks, it has been often considered as noise. Our results show it might be a rich source of physiologically relevant information. The alternative of using direct central recordings in humans to assess reflex pathways is severely limited by practical/ethical considerations \citep{mcauley2000physiological}.

Our modeling environment makes it feasible to systematically vary all physiological parameters of the system (e.g. fusimotor drive, presynaptic inhibition, muscle and load mechanical parameters, etc) to replicate experimental observations. As our capabilities to simulate neuromechanical interactions and reflex modulation becomes more complete, it may be possible to guide the design of experiments and methods to extract important information about the state of injury or dysfunction.

%% file: acknowledgement.tex
\section{Acknowledgment}
Research reported in this publication was supported by the National Institute of Arthritis and Musculoskeletal and Skin Diseases of the National Institutes of Health under Awards Number \emph{R01 AR-050520} and \emph{R01 AR-052345} to FVC. The contents of this endeavor is solely the responsibility of the authors and does not necessarily represent the official views of the National Institutes of Health. This work was also supported by the Canadian Institutes of Health Research to REK, and Fonds Qu\'eb\'ecois de la Recherche sur la Nature et les Technologies to KJ an MG.